\documentstyle[epsf,graphicx]{mn}

\def\etal{{et al.\ }}

\def\x2{$\chi^{2}$}

\def\asca{{\it ASCA }}
\def\rosat{{\it ROSAT }}

\def\bepposax{{\it BeppoSAX }}

\def\x2{$\chi^{2}$}
\def\lunits{$\rm{erg\,s^{-1}}$}
\def\funits{$\rm{erg\,s^{-1}\,cm^{-2}}$}
\def\cunits{$\rm{cm^{-2}}$}
\newbox\grsign \setbox\grsign=\hbox{$>$} \newdimen\grdimen \grdimen=\ht\grsign
\newbox\simlessbox \newbox\simgreatbox \newbox\simpropbox
\setbox\simgreatbox=\hbox{\raise.5ex\hbox{$>$}\llap
     {\lower.5ex\hbox{$\sim$}}}\ht1=\grdimen\dp1=0pt
\setbox\simlessbox=\hbox{\raise.5ex\hbox{$<$}\llap
     {\lower.5ex\hbox{$\sim$}}}\ht2=\grdimen\dp2=0pt
\setbox\simpropbox=\hbox{\raise.5ex\hbox{$\propto$}\llap
     {\lower.5ex\hbox{$\sim$}}}\ht2=\grdimen\dp2=0pt

\begin{document}

\title[ \bepposax spectrum of Mrk609] {The \bepposax spectrum of the 
composite galaxy Mrk609}

\author[A. Pappa, I. Georgantopoulos, M. Ward and A.L. Zezas ]
       {A. Pappa$^{1,2}$, I. Georgantopoulos$^{3}$, M. Ward$^{1}$,
A.L. Zezas$^{4}$ \\
$^{1}$Department of Physics and Astronomy, University of Leicester, 
Leicester, LE1 7RH, UK \\
$^{1}$Institute for Astronomy, University of Edinburgh, Royal
Observatory, Blackford Hill, Edinburgh EH9 3HJ, UK \\
$^{3}$Institute of Astronomy \& Astrophysics, 
National Observatory of Athens, Lofos Koufou, Palaia Penteli, 
15236, Athens, Greece \\
$^{4}$Harvard-Smithsonian Center for Astrophysics, 60 Gardes St., Cambridge,
MA 02138, USA \\
}

\maketitle

\label{firstpage}

\begin{abstract}
We present \bepposax observations of the starburst/Seyfert composite 
galaxy Mrk609. This enigmatic object has an optical spectrum dominated 
by the features of starburst galaxies, 
yet its X-ray luminosity  (6.3$\times10^{42}$\lunits) is typical of an AGN. 
The X-ray spectrum of Mrk609 can be parameterised by a single
power-law model with a photon index $\Gamma\sim1.6\pm0.1$
 and no evidence for significant absorption above the Galactic value. 
Long term variability in both the 0.1-2 keV and 2-10 keV energy bands is detected, 
 again suggesting that the X-ray emission is dominated by an AGN. 
 The observed broad Ha component is a factor of 40 below 
that predicted by the X-ray flux implying a deficit of 
 ionizing UV photons.    

\end{abstract}


\section{INTRODUCTION}

Moran, Halpern $\&$ Helfand (1996), after careful spectroscopy of a sample based 
on the cross-correlation of the IRAS PSC and \rosat All Sky Survey, reported the 
discovery of an ``anomalous'' class of objects. The optical spectra of these sources 
are dominated by the features of starburst galaxies, based on the
emission line diagnostic diagrams (Veilleux $\&$ Osterbrock 1987), yet their X-ray luminosities are 
typical of Seyfert 2 galaxies. 
Close examination of their optical spectra reveals some weak Seyfert-like features:
[OIII] significantly broader than all other narrow lines in the spectrum and in some cases a weak broad H$_\alpha$ 
component. The authors designated these objects ``starburst/Seyfert composite'' galaxies and 
presented them as a new class of X-ray luminous source. 
Similar "composite" objects have also been
noticed by Veron et al. (1997).
Indeed, they presented  observations of 15 objects
 with transition spectra
 ie showing the simultaneous presence of
 a strong star-forming component and an active nucleus and they showed
that fall either on the starburst region or on the borderlines 
between the different classes.
Hereafter, we will refer to these objects as composite galaxies and we 
will distinguish them from the Sy2/Starburst galaxies that show
emission from both components at all wavelengths.

%The Baldwin \etal (1981) or Veilleux $\&$ Osterbrock (1987) emission line diagnostic diagrams allows 
%secure classification of the nuclear emission line region of most galaxies into one of three categories:
%nuclear HII regions/starbursts, Seyfert 2 galaxies and LINERS.
%Composite galaxies fall either on the starburst region or on the borderlines 
%between the different classes, so that they cannot be identified unambiguously. In the literature these galaxies have 
%been termed  ``transition'' objects as well.
%Veron \etal (1997) presented spectral observations of 15 objects with ``transition'' spectra 
%at high-dispersion 
%(66$\AA$ mm$^{-1}$) around the H$_\alpha$, [NII]$\lambda\lambda$6548,6584 and/or H$_\beta$, 
%[OIII]$\lambda\lambda$4959,5007 emission lines. They showed that most of them (10) have composite spectra, 
%resulting from the simultaneous presence of a Seyfert nucleus and a
%HII region. 
%Hereafter, we will refer to these objects as composite galaxies and we 
%will distinguish them from the Sy2/Starburst galaxies that show
%emission from both components at all wavelengths.

The composite galaxies bear close resemblance to the narrow-line X-ray galaxies (NLXGs)
detected in large numbers in deep \rosat surveys (eg Boyle \etal 1995, 
Griffiths \etal 1996). These NLXGs again have spectra composite of Seyfert and 
starburst galaxies (Boyle \etal 1995) with luminosities L$_{2-10 keV} \sim 10^{42-43}$\lunits.
Unfortunately the faint fluxes of these NLXGs do not allow their detailed study in either 
optical or X-ray wavelengths.
Although it is unclear whether these nearby ``composites'' are the same class of objects as 
those found in \rosat deep field NLXGs, their high luminosities need to be explained.
It is unclear how their intense X-ray emission can be reconciled with weak or 
absent Seyfert characteristics.

\subsection{Composite galaxies in X-rays}

Only a few composite galaxies have been studied so far in X-rays.
Specifically, 
IRAS00317-2142 (Georgantopoulos 2000) has been observed with \asca and is the most luminous object
(L$_x=\sim10^{43}$\lunits in the 0.1-2 keV band) in the Moran \etal (1996) sample. 
The spectrum is represented by a power-law with $\Gamma \sim 1.76$ and 
there is no evidence for absorption above the Galactic value. 
Strong variability in the 1-2 keV band (by a factor of three) is detected between the \rosat 
and \asca observations. These characteristics indicate an AGN origin
for the X-ray emission. 
However no iron line is detected and 
the 90 per cent upper limit on the equivalent width is 0.9 keV. 
The ratio $f_{HX}/f_{[OIII]}$ $\sim$ 2.5  rule out the Compton thick interpretation 
for IRAS00317-2142. 
However, the precise nature of this object and the 
relative contribution of the starburst and AGN components could not be determined.

A further composite object studied in X-rays
with \rosat and \asca is AXJ1749+684 (Iwasawa \etal 1997). 
AXJ1749+684 was serendipitously detected with the \asca GIS. Its X-ray spectrum is flat 
($\Gamma=1.23_{-0.27}^{+0.21}$). The flatness is attributed by the
authors to 
absorption mainly because of the:
(a) large Balmer decrement in the narrow line region, $H_\alpha/H_\beta$=7.32 and 
(b) lack of significant X-ray detection at $<$0.4 keV.
On the other hand, the optical counterpart of AXJ1749+684 is detected
in the Kiso 
Schmidt Survey of 
UV-excess galaxies. 
Iwasawa \etal (1997) claimed that the UV emission is due to
large-scale starburst activity, 
however in this 
case strong far infrared emission should be expected, which is
inconsistent with the 
non-detection of this source
by IRAS. They concluded that the X-ray spectrum of AXJ1749+684 is well fitted by an obscured 
(N$_H=2.1^{+6.2}_{-2.1}\times10^{21}$ cm$^{-2}$) 
Seyfert nucleus 
embedded within a star-forming galaxy.  

Recently Levenson \etal 2001, examined NGC6221 as a further example of 
a composite galaxy. They proposed that the X-ray spectrum of this
object is characterised by a Seyfert 1 like spectrum. They detect an
iron line and continuum variability on short- and long-term
timescales. The source has a column density of N$_H$=10$^{22}$ \cunits.
They proposed that the central region is obscured by a surrounding starburst.
Thus the optical spectrum  has the characteristics of the starburst
component alone.
 
%In addition, the X-ray spectrum of the spiral galaxy NGC3628, which has been classified as a starburst galaxy,
%has a flat spectrum with $\Gamma1.16_{-0.21}^{+0.24}$.

\subsection{Mrk609}

Mrk609 is at a redshift of 0.034. The optical position of the object is 03 25 25.3, -06 08 39 (J2000) 
and the Galactic absorption is $N_H$=4.41$\times10^{20}$ cm$^{-2}$.
%It is taken from the Veron \etal (1997) sample. This sample has very good optical spectra 
%which make their classification secure.
The H$_\beta$ profile is $\sim$ 110 km sec$^{-1}$ wide, while the
[OIII] lines are $\sim$ 4 times wider. In addition, the broad
blueshifted wings seen on the [OIII] lines are completely missing in
H$_\beta$ (Heckman \etal 1981). The UV spectrum show strong
contribution from hot stars (Rudy \etal 1988) indicating the presence of an intense
starburst component in Mrk609.
 
The broad $H_\alpha/H_\beta$ value is 7.8 (Osterbrock 1981). 
In a later observation, Rudy \etal (1988) obtained a value for the broad $H_\alpha/H_\beta$=5.
The discrepancy was attributed to continuum variability.
The high broad $H_\alpha/H_\beta$ value was attributed by Osterbrock
1981 to reddening of the broad line region.
However the broad $Ly_\alpha/H_\beta$ value is 16, which is large for
Seyfert 1 galaxies, ruling out obscuration (see Rudy \etal 1988 for a
detailed discussion).

\section{Observations and Data Reduction}
Mrk609 was observed by \bepposax three times. The first observation was carried out on 
20/01/2000 for $\sim$18 ksec (LECS exposure 7.13 ksec), the second one on 14/02/2000 for $\sim$ 2.5 ksec 
(LECS exposure 1.4 ksec ) and the 
third one on 4/03/2000 for $\sim$ 28 ksec (LECS exposure 6668 ksec).
It should be reminded here, that the exposure time for the LECS is lower than the MECS because it is limited by stronger 
operational constraints to avoid UV light contamination, thus it is operated during Earth 
dark time only.
Spectra and light curves of Mrk609 have been extracted from circular regions 
centered on the source. We used a circular extraction cell of 4 and 6 arcminutes in
radius for MECS and LECS data respectively. The background spectra were extracted 
from blank deep field exposures, using the same region of the detector in each case.

In order to check whether there is any flux or spectral variability in the soft band, an unpublished \rosat
PSPC observation of Mrk609 was analysed. Mrk609 was observed by \rosat on the 29th of January 1997 for
5801 seconds. 
A source spectrum has been extracted from a circular region of $\sim$ 1.5 arcmin radius around the centroid 
of the source, while the background spectrum was extracted from an annulus of internal and external radii  
of 3 and 9 arcmin respectively. 

\section{Variability}

In order to investigate the nature of the X-ray source, we constructed 
the long term 
light curves of Mrk609.
Then the variability was tested by means of a $\chi^2$ test against the hypothesis that the flux was constant.
The $\chi^2$ values are quoted in Table 1 for four light curves: 
LECS 0.1-2 keV, 
MECS 2-10 keV, MECS 2-5 keV, MECS 5-10 keV. 
The MECS data were split in three different bands in order to check whether there are different components 
contributing to the variability of 
the source. Fig. 1 shows the light curves in the four energy bands.
The points are in chronological order; 
The errors are based on counting statistics only.

It is evident that Mrk609 is variable in both soft and hard energies.
In the LECS the flux is decreasing and the difference in flux between the first and third observations
is about a factor of 2.
Significant variability at the $>$99.9 per cent confidence level is detected in the 2-10 keV 
and 2-5 keV energy bands, whereas the variability in the 5-10 keV band is significant only at
the 57 per cent confidence level. 
In the three hard bands the variability follows the same trend; the first and third observations have 
comparable fluxes, whereas during the second observation the flux drops by up to a factor of 3.
This suggests that the whole hard X-ray emission is produced by the same mechanism and in the same region.
However, the variability in the soft band follows a different trend
from the hard band, suggesting that emission in the two bands is
produced by different mechanisms and or in different regions.

%The presence of variability rather argues against an HII origin for
%the emission. However see Chandra observation of M82, where strong
%variability is observed. In M82 the variability is attributed solely
%to X-ray binaries (Kaaret \etal 2000).

We also checked for short time variability. Mrk609 does not show
evidence for short time variability at a significant level. This is
probably due to the low statistics.

\begin{table}
\begin{center}
\caption{$\chi^2$ for the long-term light curves.}

\begin{tabular}{lcc} 
\hline 
Energy  & $\chi^2$(dof) & null hypothesis \\
band            &           &     probability  \\ \hline

0.1-2 keV & 8.16(2)& 1.7$\times10^{-2}$\\
2-5 keV & 30.08(2)&2.9$\times10^{-7}$\\
5-10 keV & 2.76(2) & 0.43\\
2-10 keV & 45(2)&1.4$\times10^{-10}$   \\ \hline

\end{tabular}
\end{center}
\end{table}

\begin{figure*}
\begin{center}
\begin{tabular}{ll}
{\large {(a)}} &  {\large {(b)}} \\
\rotatebox{0}{\includegraphics[height=8.0cm]{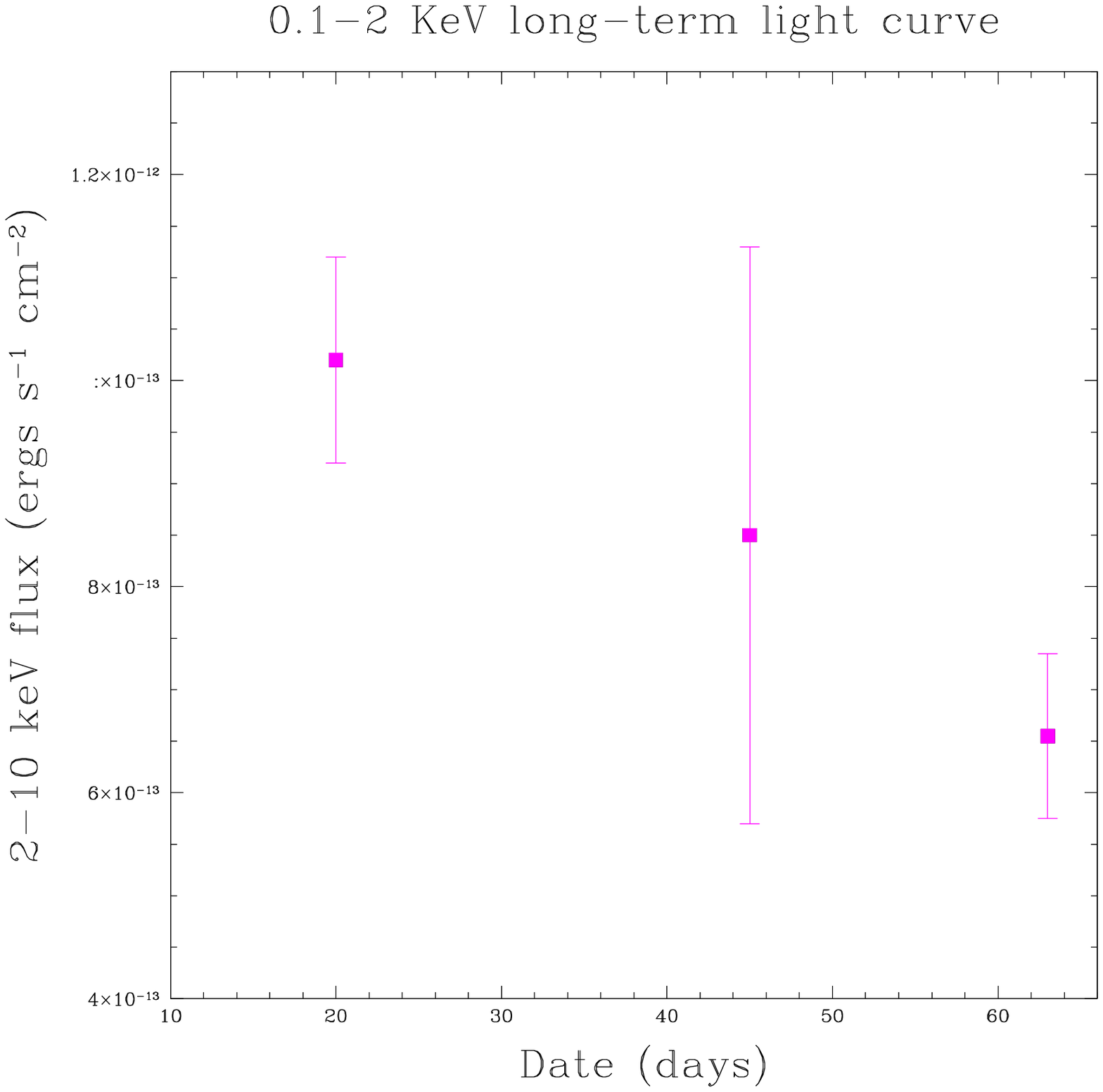}}
 &\rotatebox{0}{\includegraphics[height=8.0cm]{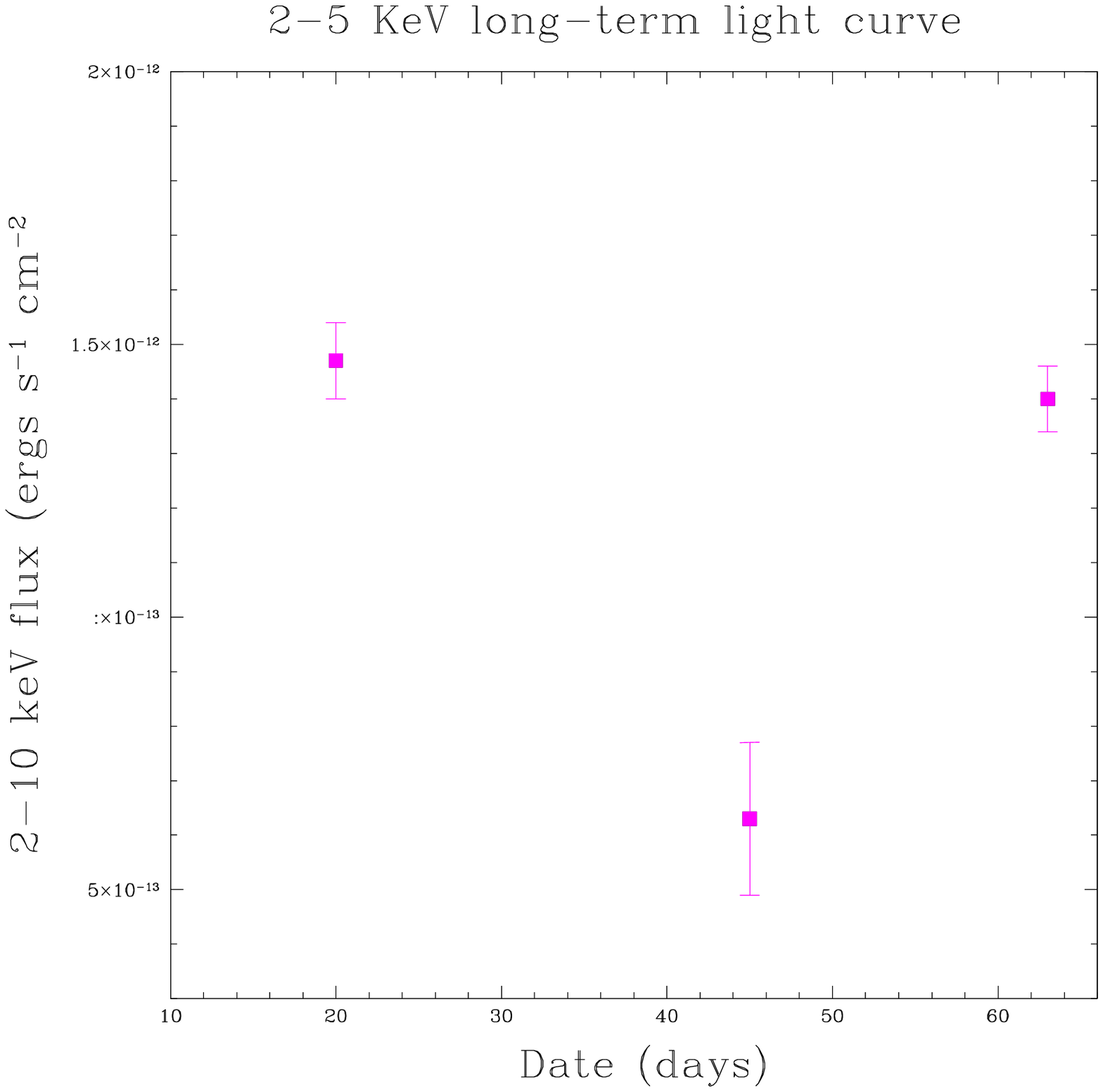}}\\

{\large {(c)}} & {\large {(d)}}  \\
\rotatebox{0}{\includegraphics[height=8.0cm]{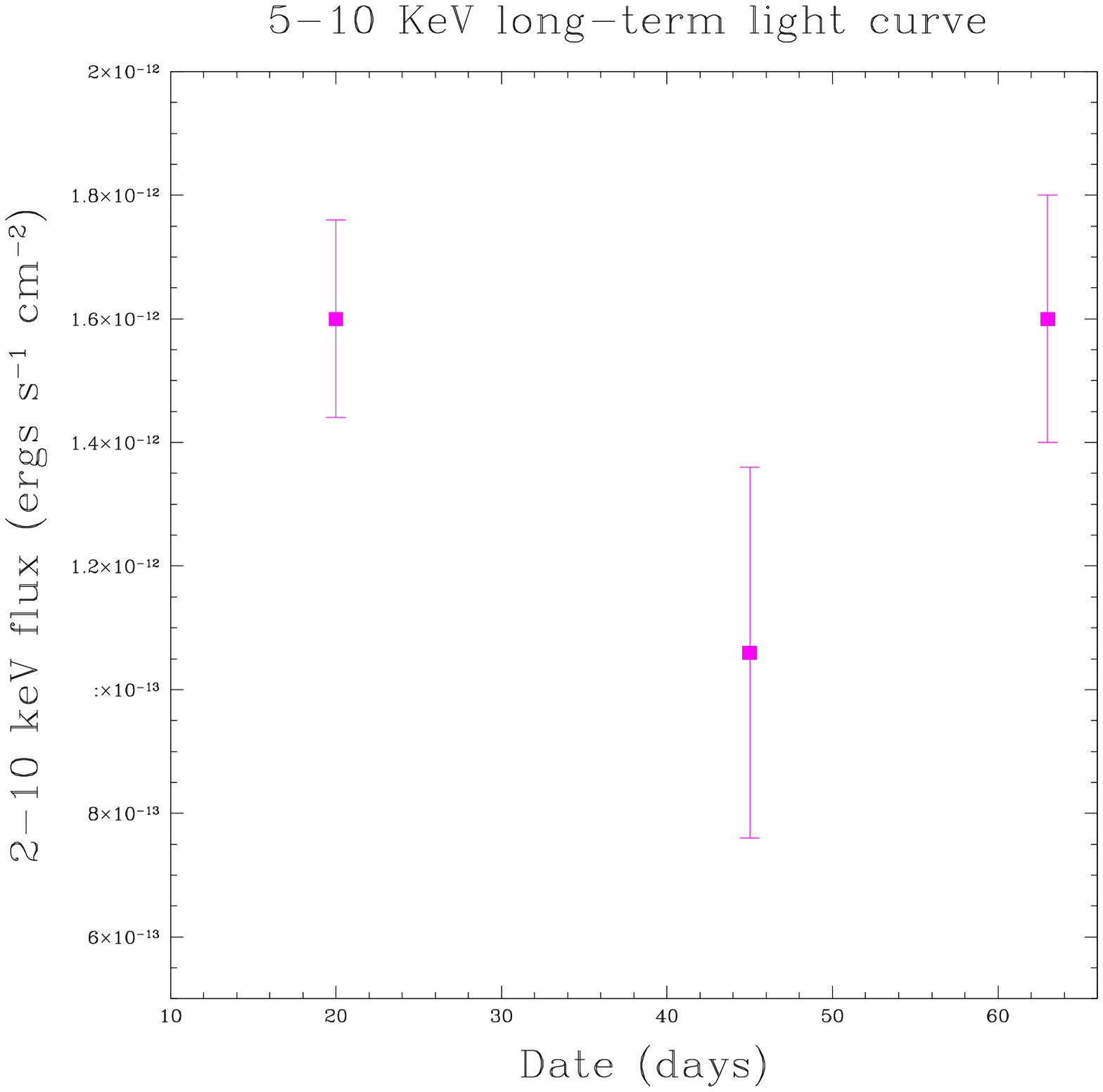}}
&
\rotatebox{0}{\includegraphics[height=8.0cm]{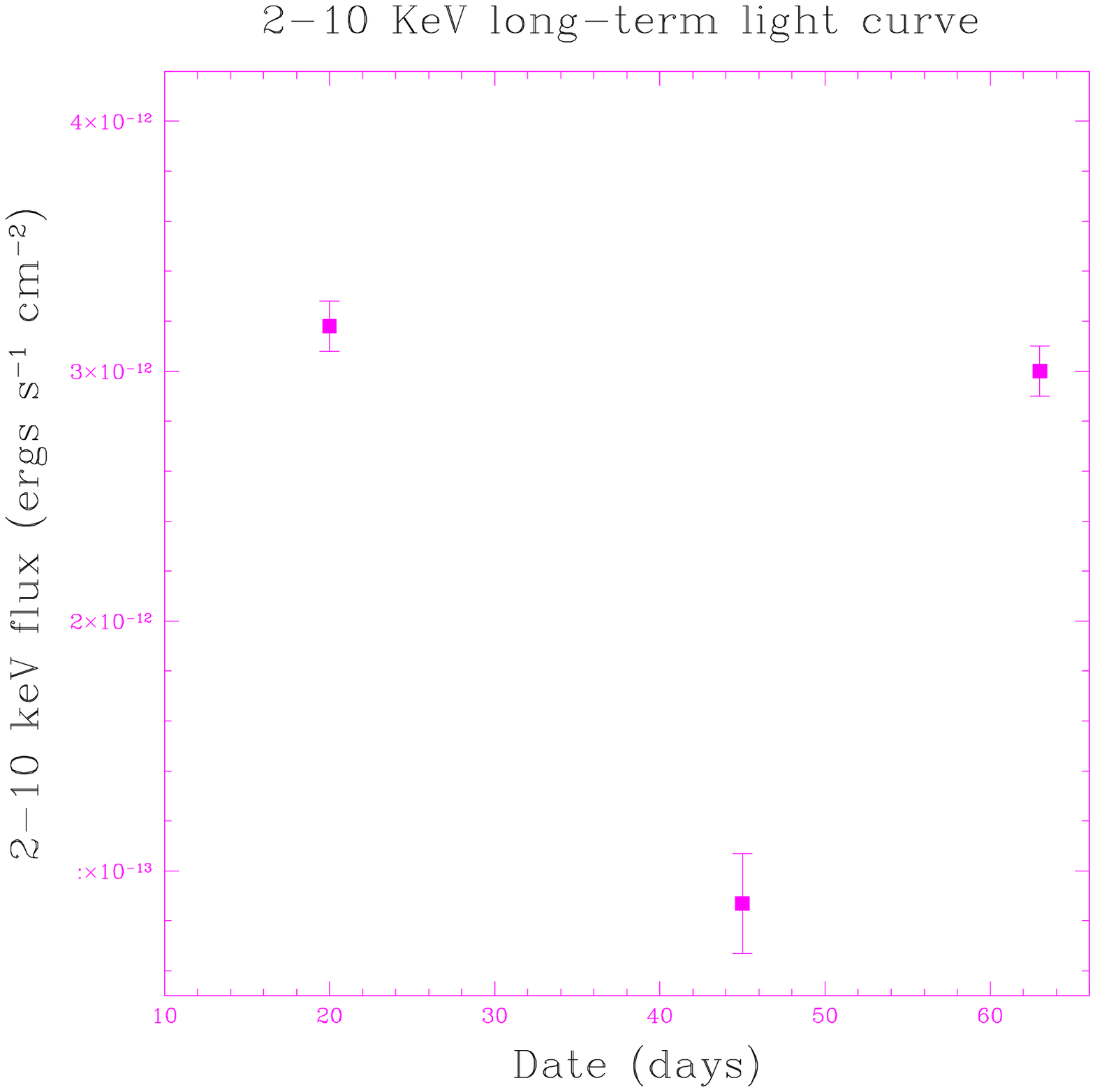}} \\

\end{tabular}
\end{center}

\caption{The long term light curves for Mrk609; (a) the 0.1-2 keV, (b) 
2-5 keV, (c) 5-10 keV, (d) 2-10 keV curves.} 
\end{figure*}

\section{SPECTRAL ANALYSIS}

The spectral fitting was performed in the 0.12-10 keV band. The appropriate redistributions matrix file 
and the ancillary response file for the observation date were obtained from the \bepposax Science Data Center 
archive. A summary of the spectral fitting results is given in Table 2. In the following analysis all 
three \bepposax observations are fitted together.

Throughout this paper values of H$_o$=75 km s$^{-1}$ Mpc$^{-1}$ and
q$_o$=0.5 are assumed.

\subsection{The AGN models} 

We first fit the data with a single power-law model (PL). We obtain an acceptable fit ($\chi^2$=98.89 for 82 dof) 
with $\Gamma=1.57_{-0.10}^{+0.10}$, and N$_H \leq$ 1.32$\times$
10$^{21}$\cunits.
This model together with the data points and the data to model ratio are
 plotted in Figure 2. 
The observed flux in the 2-10 keV band for this model is 2.86$\times10^{-12}$ \funits, which corresponds to a 
luminosity of 6.3$\times10^{42}$ \lunits in the same band.
If the slope of the power-law model is fixed at the 1.9 value, the nominal value for the Seyfert 1 galaxies (
Nandra $\&$ Pounds 1994), the model yields an unacceptable fit ($\chi^2$=119.26 for 83 dof) with 
$N_H=2.03_{-1.8}^{+1.5} \times 10^{21}$ cm$^{-2}$. 

Although Seyfert galaxies show strong narrow iron K$_\alpha$ emission lines, no
such line is detected in Mrk609. However an upper 
limit of $\sim283$ eV is obtained, consistent with values seen in
Seyfert galaxies.

\begin{figure*}
\rotatebox{270}{\includegraphics[height=11.5cm]{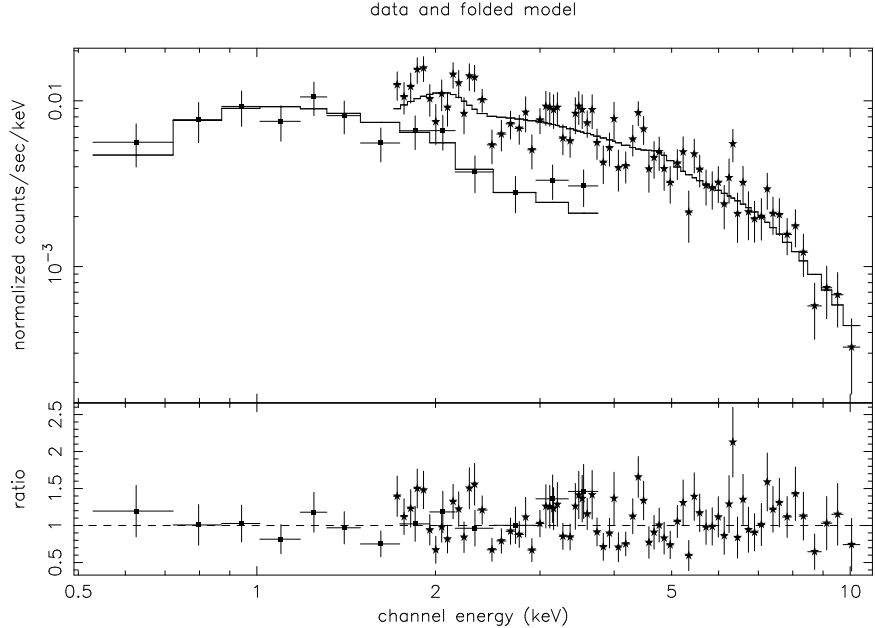}}
\caption{The \bepposax time averaged spectrum, when the
single power-law model is applied to the data. The filled squares
represent the LECS data points and the stars the MECS data points. The 
top panel shows the data with the model and the bottom panel shows the 
data/model ratio}
\end{figure*}

In the context of the unified models we expect to see some faction of the primary emission through the torus, 
with a component superimposed upon this that represents a fraction of emission scattered back into our line 
of sight by material lying above the torus. This model is representing by two power-laws 
with the same
photon index but different normalisations and absorptions - the scattering model. 
When this model is applied to the data, the normalisations
of the two power-laws are comparable, whereas no excess absorption above the Galactic is required
 and thus this model effectively is the same as the single power-law model.
Therefore it is evident that the scattering model does not provide a good representation of the data.
%In addition and in the context of the unified model, the source may be partially covered by the obscuring
%medium, so that a small fraction of the primary emission escapes, while the rest is reprocessed by the torus.
%Thus a partial covering model has been applied to the data. 
%Observations in the optical shows that the direct emission is only a few per cent of the emission passing through 
%the torus. Because of this and due to the low quality of the data, particularly at lower energies the covering 
%fraction is fixed at the 0.9. The model yields an acceptable fit 

Finally, an ionised warm absorber model in addition to the Galactic column density was fitted to the data (PL+warm).
The temperature of the absorber is fixed at T=10$^5$ K (Brandt \etal 1999). The model provides an acceptable 
fit ($\chi^2$=98.84 for 81 dof) but does not represent a statistically significant improvement to the single
power-law model. The best fit parameters are 
$\Gamma=1.60_{-0.12}^{+0.16}$
and warm column density $N_{H_w}=6.73_{-6.73}^{+25.47} \times 10^{21}$ cm$^{-2}$, while the
ionisation parameter is practically unconstrained, possibly 
because of the poor statistics of the data.

\begin{table*}
\begin{center}
\caption{The spectral fits results on the \bepposax data.}

\begin{tabular}{lcccccc} 
\hline 
Model  & $\Gamma$ & $N_H^{\, a}$ & $N_{H_w}^{\, b}$ & $\xi$ & kT$_{RS}$& $\chi^2$(
dof) \\ \hline
single PL &$1.57_{-0.10}^{+0.10}$ &$\leq$1.32&-&-&-& 98.89(82)  \\
          &1.9f&$2.03_{-0.98}^{+1.53}$&&&&119.26(83) \\
PL+warm   &$1.60_{-0.12}^{+0.16}$  & g & $\leq$369.11 & $\geq0$ &-&97
.90(81)\\
PL+RS     &$1.57_{-0.11}^{+0.09}$ &g&-&-&0.8f& 99.00(81)\\
%PL+RS     &$2.81_{-0.96}^{+2.59}$ &g&-&-&$\geq27.87$& 95.0(80)\\
RS        &-&g&-&-&18.92$_{-6.02}^{+10.05}$ &108.94(83)\\
2RS       &-&g&-&-&$\leq1$ & 105.57(80)\\
          &&&&& 21.81$_{-7.69}^{+14.33}$ & \\ \hline

\end{tabular}
\end{center}
NOTE: g indicates that the $N_H$ is set to the Galactic value.\\
 
$^a$ column density in units of $10^{21}$\cunits. \\
$^b$ column density in the warm medium in units of 10$^{21}$\cunits. \\
\end{table*}

\subsection{The composite model}

Given the composite classification of this object, 
it is natural to investigate a model in which 
X-ray emission originates from both a starburst and an AGN. 
A power-law plus a Raymond-Smith model (PL+RS) with the temperature fixed 
at 0.8 keV is adopted.
The power-law component is allowed to have additional
absorption over and above that of the thermal component. This model yields an acceptable fit
($\chi^2$=99.81 for 81 dof) with 
$\Gamma=1.57_{-0.11}^{+0.09}$, whereas no excess absorption above the Galactic is required.
When the temperature of the thermal component is a free parameter, we find kT $> $18 keV
and $\Gamma=2.81_{-0.96}^{+2.59}$ ($\chi^2$=95.00 for 80 dof).
However, the temperature of the thermal component is far too high for a starburst and thus this model cannot provide a
physical description of the data.

\subsection{The pure starburst model}

For completeness we have also investigated 
pure starburst models.
Firstly a single Raymond-Smith model (RS) was fitted to the data. An acceptable fit was obtained 
($\chi^2$=108.94 for 83 dof) with kT$_1$=18.92$_{-6.02}^{+10.15}$ keV. 
No starbursts with such a high temperature have been found as yet. 

Then a two Raymond-Smith model representing thermal emission from a 
pure starburst galaxy following Zezas \etal (1998) was utilised.
The soft emission is parameterised by a thermal component with kT$_1
\leq$ 1 keV and the emission in the hard band by
kT$_2$=21.8$_{-7.69}^{+14.33}$ keV, again too high for a starburst. 
Therefore it is obvious that this model cannot provide a physical description
of the data.

%The model yields an acceptable fit ($\chi^2$=109.94 for 80 dof), 
%however the model does not provide a physical description of the data with  
%kT$_1$=21.8$_{-7.69}^{+14.33}$ keV to represent the soft emission and kT$_2 \geq $25.45 to represent the hard 
%emission, again too high for a starburst. 

From the above it is evident that the pure starburst model is ruled out, whereas the composite model does 
not provide 
a physically accepted model.
% apart from the case where the temperature is set to kT=0.8 keV. In addition 
%the single power-law model yields an acceptable fit.

\section{Spectral variability}

Mrk609 was observed three times with \bepposax, allowing us to examine whether there is spectral variability.
In particular it is interesting to see whether the drop in the flux during the second observation is related 
to a change in the spectrum of Mrk609 and examine whether the X-ray behaviour is similar to that  
of black hole candidates (BHC) in our galaxy during their high and low states.
For this analysis only the single power-law is applied to the MECS data. 

In addition any spectral variability at soft energies will be examined
by comparing the \rosat PSPC and LECS data. For that reason we
analysed an unpublished observation of Mrk609 with \rosat as explained 
in section 2.

\subsection{Observation 1}

The power-law model yields a fit
($\chi^2$=55.75 for 38 dof) with $\Gamma=1.63_{-0.16}^{+0.16}$ and Galactic absorption.
The observed flux in the 2-10 keV band for this model is 2.85$\times10^{-12}$ \funits, which corresponds to a 
luminosity of 6.32$\times10^{42}$ \lunits in the same band.
%The model with the data and the data to model ratio is plotted in Figure 7.23.
We do not detect an iron line at 6.4 keV but we obtain an upper limit
of 540 eV.

%\begin{figure*}
%\rotatebox{270}{\includegraphics[height=11.5cm]{obs1_nomanual_po.ps}}
%\caption{The \bepposax spectrum for the first observation, when the
%single power-law model is applied to the data. The filled squares
%represent the LECS data points and the stars the MECS data points. The 
%top panel shows the data with the model and the bottom panel shows the 
%data/model ratio.}
%\end{figure*}

\subsection{Observation 2}

During the second observation Mrk609 was observed for $\sim$ 2.5 ksec
only, and no reliable spectrum could be
extracted. However clues for the spectral shape of the spectrum of Mrk609 during this short observation
come from the hardness ratio (HR) of the source. Here the HR is defined as h-s/h+s, where h 
and s are the total number of counts, in the 2-10 keV and 1-2 keV respectively. Only MECS data were used since
Mrk609 is not detected in the LECS. The HR is 0.31$\pm$0.18, which corresponds to a power-law of 
$\Gamma=1.2\pm0.70$ assuming Galactic absorption. Unfortunately the uncertainty in the value of $\Gamma$ is
too high for any firm conclusion about spectral changes to be derived.

\subsection{Observation 3}

The power-law model yields a $\Gamma=1.64_{-0.14}^{+0.14}$ and Galactic
absorption ($\chi^2$=61.16 for 48 dof).
The observed flux in the 2-10 keV band for this model is 2.72$\times10^{-12}$ \funits, which corresponds to a 
luminosity of 6.02$\times10^{42}$ \lunits in the same band.
%The model with the data and the data to model ratio is plotted in Figure 7.24.
In the energy range 6-7 keV there is some evidence for residuals. So a 
Gaussian line was added to the model. The best fit energy for the line 
is 6.72$_{-0.24}^{+0.22}$ keV, clearly inconsistent with the line
originating from cold iron.
The model is acceptable ($\chi^2$=56.31 for 46 dof) with $\Gamma=1.70_{-0.15}^{+0.16}$ and an equivalent width 
for the iron line 394$_{-286}^{+391}$ eV.
%However,the inclusion of the iron line does not improve the fit 
%significantly (only at the 80 per cent confidence level). 
%The power-law plus the 6.4 keV line model with the data and the data to model ratio 
%is plotted in Figure 7.25.
%To illustrate any spectral variability and in particular to further
%investigate the presence of an emission line 
%in the 6-7 keV energy range, the third to first observation ratio against the energy bins is plotted (Figure 7.26). The% figure clearly shows that indeed there is an excess of counts in the energy range 6-7 keV during the third observation% and confirms the detection of an iron line during the third observation. 
%It is noted the continuum did not vary between these two observations.  

\subsection{LECS/PSPC spectra comparison}

The \rosat spectrum is well represented by a single power-law with
$\Gamma=2.02_{-0.25}^{+0.24}$, steeper 
than 
the \bepposax spectrum ($\Gamma \sim 1.6$). 
However due to the poor statistics of the LECS data compared to the MECS, the spectral slope of the former
is probably driven by the latter and thus the discrepancy might not be real.
Therefore, in order to check whether the PSPC and LECS slopes are
inconsistent, we fit 
the LECS data alone with a single power-law model in the 0.12-2.0 keV energy range.
The best fit parameters are $\Gamma=1.97_{-0.53}^{+0.51}$ and Galactic absorption 
($\chi^2$=3.22 for 6 dof). Due to error uncertainties the slope is
consistent with both the PSPC and MECS slopes, 
and thus any variability between the \rosat and \bepposax observations as well as any spectral upturn within the
\bepposax observations cannot be examined.

\section{Discussion}

Mrk609 displays an X-ray continuum, which is somewhat at odds with its
optically composite nature.
A simple power-law is a good description of the data over the whole
0.1-10 keV energy range, 
with a rather flat index 
($\Gamma=1.57^{+0.10}_{-0.10}$). 
 Still, this value is not inconsistent 
 with those encountered in Seyfert-1 spectra
 (Nandra \& Pounds 1994).  
Moreover no iron emission line was
detected when all three 
observations are fitted together and 
the upper limit of the 
equivalent width of the line is $\sim$ 283 eV.
For the third observation, where the data-to-model ratio shows line
like residuals in 
the 6-7 keV band,
the inclusion of the iron line improves the fit significantly. The
best fit line is 6.72$_{-0.24}^{+0.22}$ keV with an equivalent width
of 394$_{-286}^{+391}$ eV.
Nevertheless the detection of significant X-ray variability confirms
that a super-massive black hole resides in Mrk609 and powers 
the X-ray emission, and rules out significant contribution from a
starburst component to the X-ray emission.

The data do not allow us to constrain the X-ray column density and we
obtain a 90 per cent upper limit of $1.3\times10^{22}$\cunits.
However, optical and ultraviolet observations are in favour of low
obscuration towards the central engine.
%We find no significant evidence for the presence of an obscuring column.
%However, the uncertainties remain large: 
% the 90 per cent upper limit is rather high $1.3\times10^{22}$\cunits.
%The absence of an X-ray obscuring column is 
% in agreement with  optical and UV observations. 
 Indeed, Rudy et al. (1988) find 
%find a small Balmer decrement together with 
a high $Ly_\alpha/H_\alpha$ ratio,
 implying that the optical redenning is negligible.  

Recently Levenson \etal 2001, proposed an 'obscuring starburst model' 
to explain the multiwavelength properties of the composite galaxy
 NGC6221. According to
their model NGC6221 is a Seyfert 1 galaxy which is surrounded by a
starburst component. The starburst accounts for the X-ray obscuration
(N$_H \sim$ 10$^{22}$) \cunits and its characteristics dominate 
the optical spectrum.
Although in principal this model can explain qualitatively the 
optical appearance of the composite galaxies it doesn't seem to 
fit the X-ray observations of Mrk609. Our object does not show
concrete evidence for significant X-ray absorption. In addition the
soft X-ray variability and the high luminosity at low energies
(L$_{0.5-2 keV}\sim$2$\times10^{42}$\lunits) probably rule out a dusty obscuring 
circumnuclear starburst.
We note here that the spectral X-ray properties 
of Mrk609 are similar to the composite IRAS00317-2142
(Georgantopoulos \etal 2000). Again this galaxy has a low column
density, consistent with the Galactic, and thus the obscuring
starburst model cannot explain the properties of IRAS00317-2142.

Although the single power-law model yields a good representation of the Mrk609 spectrum the X-ray 
long term variability indicates that the spectrum of Mrk609 consists
of more than one components.
An AGN covered by a warm absorbing screen could provide an explanation for the observed long term variability.
In this case changes in the X-ray continuum flux, will be followed by changes in the ionisation state of the 
warm absorber resulting in changes in the emission in the soft band.
However, the quality of the data does not allow to examine the
viability of the model to Mrk609.

Given the composite nature of Mrk609 it is natural to investigate
whether emission from starburst regions contribute to the X-ray wavelengths.
In principle, in a composite starburst-AGN model, the power-law component is heavily absorbed, 
and thus the star-forming component, which is 
located outside the obscuring screen, dominates the soft emission.
However, when this model is applied to Mrk609 data, no excess absorption above the Galactic 
is required by the data for the 
power-law component. In addition the poor quality of the data at
energies below $\sim$2 keV do not allow us to constrain the
temperature of the thermal emission and make an unambiguous estimate
of the starburst contribution to the X-ray emission.
The strength of the star-forming component 
 can be indirectly estimated from the observed IR flux. 
The expected X-ray contribution from stars was calculated using 
 the empirical relationship between 
infrared and X-ray luminosity (equation [2] David, Jones $\&$
Forman 1992)
 found in a sample of IRAS galaxies.
 However, note that some of the infrared (IR) flux could arise from nuclear
 reprocessed emission from the obscuring medium. 
 Thus any starburst contribution to the X-ray flux derived  
 using the above relation may be
overestimated and the derived flux  should only be treated as an 
upper limit. 
First we calculated the IR luminosities using the fluxes IRAS fluxes
at 60${\mu}m$ and 100${\mu}m$ and equation [1] from David, Jones \etal
1992.
We find that the upper limit of the expected contribution of a starburst in the 0.5-4 keV band is 
2.75$\times10^{42}$\lunits. The luminosity in the same band derived by the spectral fitting is 
$\sim$4.7$\times10^{42}$ \lunits, indicating that about half of the
soft emission may be due to an intense starburst component. 
However, the variable soft X-ray emission clearly argues  that any starburst 
contribution  in soft energy band  should be low.

To further test the AGN interpretation of Mrk609 the broad H$_\alpha$ line and the 2-10 keV flux were compared. 
Ward \etal (1988) found a strong correlation between the two
quantities in a sample of IRAS selected Seyfert 1 galaxies. 
The observed luminosity of the broad H$_\alpha$ is L(H$_\alpha$)=8.4$\times10^{39}$\lunits, 
 whereas the 2-10 keV luminosity is $\sim$7.5$\times10^{42}$\lunits. 
According to the above relation the predicted broad H$_\alpha$ luminosity should be $\sim$40 times higher. 
The above discrepancy between the optical and X-ray spectrum could be explained by variability. 
Possibly the AGN was weaker during the optical observation (1984), but
brightened over the $\sim$15 years timescale between the optical and
X-ray observations. 
 Alternatively the source may have unusually low UV emission. Then the photoionised emission lines 
would have lower fluxes than those typical for AGNs. 
To examine this possibility the Mrk609 spectral energy distribution
(SED) was computed. This is shown in Figure 3. It is indeed clear that
Mrk609 lacks a big blue bump (BBB). 
This feature is characteristic of high luminosity unobscured AGNs and is thought
to be a signature of the presence of a cold accretion disk around the
black hole (see Koratkar $\&$ Blaes 1999). 
We note here that lack of ultraviolet excess has also been observed in a sample of
low-luminosity AGN (Ho 1999). Low accretion rate models have been
employed to account for the absence of the BBB feature. 
Note that Mrk609 has a very strong $Ly_\alpha$ line
 (Rudy et al. 1988). 
%The only known objects with stronger Ly$_\alpha/H_\beta$ 
% ratios are the Seyfert-1 galaxies Mrk359 and Mrk1018.
 The abnormally strong $Ly_\alpha$ line and 
anomalous emission line strengths in Mrk609 could be explained if the 
optical depth and ionisation parameter in the region where 
 the lines form is significantly 
 less than believed typical for Seyfert-1 galaxies.
In this scenario the discrepancy between the optical and X-ray
spectrum could be explained. 
%According to Rudy et al. (1988) the abnormally strong $Ly_\alpha$
%line is readily explained 
% if the Lyman continuum optical depth in the region where 
% the lines form is significantly 
% less than believed typical for Seyfert-1 galaxies. 
%In particular, low optical depths and ionisation parameters can produce
%large broad-line decrements and anomalous emission line strenghts.  
Using Pa$\beta$ spectroscopy Goodrich (1990) and Rix \etal (1990)
showed that Mrk609 line properties are indeed well explained by the
optical depth/ionisation parameter theory.

\begin{figure*}
\rotatebox{270}{\includegraphics[height=11.5cm]{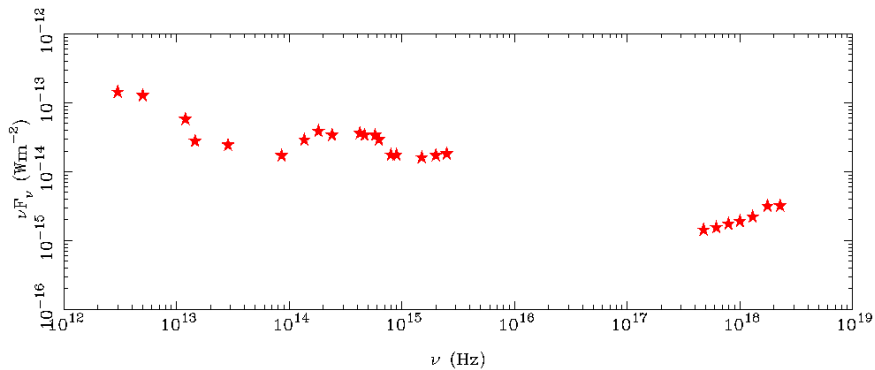}}
\caption{The spectral energy distribution of Mrk609, from far infrared 
to hard X-rays.}
\end{figure*}

As mentioned in the introduction the optical spectrum of the composite 
objects like Mrk609 bear close 
resemblance to the narrow line X-ray galaxies detected in \rosat surveys.
The X-ray spectrum of these sources is flat (Almaini \etal 1996) but
it is unclear whether the flatness of the spectrum is intrinsic or due 
to obscuration.
On the other hand, our observations show that Mrk609  has a relatively
steep X-ray spectrum and no significant X-ray absorption.
If the narrow line X-ray galaxies detected in \rosat surveys have
X-ray spectra similar to Mrk609 then they should not contribute
significantly to the XRB.

\section{summary}

We have analysed \bepposax data of the composite galaxy Mrk609.
The spectrum is described by a power-law
 $\Gamma=1.6$ with negligible absorption. 
 The absence of absorption is consistent 
 with the small Balmer decrement 
 and the large $Ly_a$ flux observed (Rudy et al. 1988). 
 The absence of an obscuring column clearly does not fit 
 the absorbed starburst model
 proposed by Levenson \etal 2001 to explain the multiwavelength
 properties of the composite galaxy NGC6221.
The detection of significant soft and
 hard X-ray variability, clearly suggests that
the AGN emission dominates the X-ray spectrum. 
Any starburst contribution to the X-ray emission should be small.
In addition Mrk609 does not follow the 
 $L_{H_\alpha}-L_x$ correlation of bright AGN (Ward \etal 1988), showing a
 weak broad H$_\alpha$ component ($\sim$40 times 
less than predicted by the X-ray flux). The discrepancy between the optical and 
X-ray spectrum can be explained as a deficit of UV ionising photons. 
 This is supported by the SED, which shows no 
 upturn of the spectrum below $3000 A$,
 implying the absence of a UV bump.  
 Alternatively, as the optical and the X-ray observations were 
 taken 15 years apart, dramatic variability in the X-ray flux could 
  result in a low  $L_{H_\alpha}-L_x$ ratio.  Finally the above
discrepancy and anomalous line properties could be explained by small
optical depth and ionisation parameter in the line emitting regions.

\section{Acknowledgments}

We would like to thank the referee Dr. J. Halpern for useful
comments and suggestions and A. Burston for producing the SED of Mrk609.

\end{document}